# QUANTITATIVE ANALYSIS OF A FRACTURE SURFACE BY ATOMIC FORCE MICROSCOPY

Pascal DAGUIER*, Stéphane HENAUX†, Elisabeth BOUCHAUD*, François CREUZET†

* O.N.E.R.A. (OM), 29 Avenue de la Division Leclerc
B.P. 72, 92322 Châtillon Cedex, FRANCE

† Laboratoire CNRS/Saint-Gobain "Surface du verre et interfaces"
39, Quai Lucien Lefranc
B.P. 135, 93303 Aubervilliers Cedex, FRANCE

<u>Abstract</u>: The fracture surface of a $Ti_3Al$-based alloy is studied using both an atomic force microscope and a standard scanning electron microscope. Results are shown to be *quantitatively comparable*. Two fracture regimes are observed. It is shown in particular that the roughness index characterizing the small lengthscales regime is equal to 0.5. Furthermore, the large lengthscales fractal domain is found to spread over nearly *six decades* of lengthscales.

## I. INTRODUCTION

Since the pioneering work of B. Mandelbrot and coworkers[1], it has been shown on all sorts of materials (steels [1, 2], aluminium alloys [6], rocks [3], intermetallic compounds [4, 5], ceramics [7]), using various experimental techniques, that fracture surfaces are self-affine and exhibit scaling properties on two [4, 5] or three decades [6] of lengthscales. In most cases, the roughness index $\zeta$ is found to lie around the value 0.8, and it was suggested that this could well be a *universal* value, i.e. independant of the material and of the fracture mode [6]. As a matter of fact, it is now believed that fracture toughness is not correlated to $\zeta$, but rather to relevant lengthscales measured on the fracture surface [9].

As far as metallic materials are considered, however, significantly smaller exponents are determined through STM (Scanning Tunneling Microscopy) experiments, i.e. for lengthscales lying in the nanometer range. Values of the roughness index close to 0.6 for fractured tungstene (regular stepped region), or to 0.5 for graphite, are reported by Milmann and coworkers [10, 11]. On the other hand, low cycle fatigue experiments on a steel sample have led to a value of $\zeta$ close to 0.6 [12]. More recently, it was shown that a new small lengthscales index could indeed be seen with standard scanning electron microscopy (SEM) [8], which was associated to a "quasi-static" fracture regime. However, this small lengthscales roughness index, lying between 0.4 and 0.5, could not be determined very precisely. It is one of the scopes of the Atomic Force Microscopy (AFM) experiments to improve the precision on this exponent.

The upsurge of interest in the problem of crack propagation through brittle heterogeneous materials, combined with the progress made in statistical physics in the understanding of line pinning by randomly distributed impurities has led very recently to a few interesting models. In fact, it was proposed that the fracture surface could be modelled as the *trace* left behind by a line (the crack front) moving through randomly distributed microstructural obstacles. The crack front motion is described by a local non linear Langevin equation [13] first written by Ertas and Kardar in a very different context [14, 15, 16, 17]. This equation describes a large number of regimes, depending on the relative values of the prefactors of the non linear terms. Although this first model suffers from some weaknesses [25], it suggests the existence of two fracture regimes. For crack velocities tending to zero, i.e. in the vicinity of the so-called "depinning transition" in the line trapping problem, the roughness index $\zeta_\perp$ perpendicular to the direction of propagation of the crack is predicted to be equal to 0.5. In the dynamic case, this exponent is predicted to be also 0.5, except in a small region of the parameters space (these parameters being essentially the prefactors of the non linearities), where $\zeta_\perp = 0.75$. More generally, for a given velocity, there is a crossover at some lengthscale from the "quasi-static" to the "dynamic" behaviour ; the cross-over length decreases quite rapidly with the crack velocity (velocity to the power $\phi = -3$). Correlatively, it is expected, within the framework of this model, that the crossover length decreases with increasing stress intensity factor $K_I$.

Another interesting model was proposed by S. Roux *et al.* [18, 19] for the fracture of plastic materials: the fracture surface is expected then to be a *minimum surface* [20], the roughness index of which is lying between 0.4 and 0.5 [21, 22, 23, 24]. Note that this mechanism, although different in its physical content than the "line depinning" one described above, is also "quasi-static", since it is based upon an equilibrium model. In this case, however, the crossover between the small and the large lengthscales regimes should lie around the plastic zone size, and hence, *increase* with $K_I$. We are faced then with two different models, which both predict a small lengthscales regime characterised by a roughness index 0.5, but in one case the crossover length is predicted to decrease with the stress intensity factor, and in the second case, it increases with that parameter as the plastic zone size.

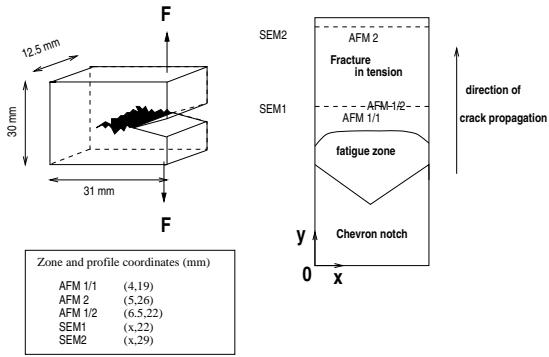

Fig. 1. Sketch of the studied samples.
**1a-** Crack propagation in mode I in a compact tension specimen.
**1b-** Sketch of the broken sample: profiles SEM1 and SEM2 observed with a standard SEM on one of the two fracture surfaces are located respectively close to zones AFM1/1,2 and AFM2, observed on the other fracture surface with an AFM.

In this paper, we report two series of experiments performed with a SEM and with an AFM : two regions on a fracture surface, corresponding to two different stress intensity factors and, very likely, to different crack velocities (although those could not be measured) are analysed on the same specimen. The results obtained with the SEM are shown to be *quantitatively comparable* to those obtained with an AFM. It is shown that the small lengthscales/"quasi-static" exponent is indeed very close to 0.5, and that the crossover length between this regime and the large lengthscales/"dynamic" regime decreases with the distance to the initial notch, thus confirming previous results [8]. In the region far from the notch, the scaling domain is found to spread over nearly *six decades* in lengthscales (5nm-1mm).

## II. EXPERIMENTAL

A notched Compact Tension specimen (see Fig.1a) of the Super-$\alpha_2$ alloy (Ti$_3$Al-based) is precracked in fatigue. Fracture is achieved through uniaxial tension (mode I) with a constant opening rate (0.2 mm/mn). The microstructure of our material is mainly constituted of $\alpha_2$ needles ($\simeq 1\mu$m thick and $\simeq 20\mu$m long) which break in cleavage, within a $\beta$ matrix, the plastic behaviour of which was shown to be important as far as the alloy fracture toughness is concerned. One of the two surfaces obtained is electrochemically Ni-Pd plated for SEM observations, while the other one is used for AFM.

Two profiles (SEM1 and SEM2, see Fig.1b) located respectively just behind the fatigue zone and closer to the edge of the specimen, are obtained by subsequently cutting and polishing the sample perpendicularly to the direction of crack propagation. These profiles are observed with a scanning electron microscope Zeiss DSM 960 at various magnifications, ranging from x50 to x3000 or x10 000, with a backscattered electron contrast. Images in 256 grey levels are registred and the profiles are extracted by image analysis (Visilog 4.1.1). The length of the images is 1024 pixels, and adjacent fields (overlapping over fifty pixels with each other) have been explored in order to build up profiles of 6 000 and 7 000 points.

Ten profiles are registred in each of the three different regions, AFM1/1 and AFM1/2, comparable to SEM1, and AFM2, comparable to SEM2 (see Fig. 1b). The lengths the profiles are 2.5, 20 and 1 $\mu$m, in regions AFM1/1, AFM1/2 and AFM2 respectively, with 10 000 points for profiles registered in zones AFM1/1 and AFM2, and 20 000 points for those in AFM1/2.

As it can be seen in Fig. 1b, there might be a slight difference of localisation between SEM1 and AFM1/1 or AFM1/2, and between SEM2 and AFM2, lying within the millimeter range. However, the experimental points registered in regions AFM1/1-AFM1/2 and SEM1 on one hand, and AFM2-SEM2 on the other hand, nicely collapse on the same curve for all the statistical analysis performed (see below). Hence, as it is expected within the framework of one or the other of the two models quoted previously, the stress intensity factor is the relevant parameter, they do not vary much within the considered regions. Note that the overlap region of the two techniques extends over two decades (see the inserts of Fig. 2 to 4).

In order to determine the roughness exponent and the cross-over length, three methods are used: the 'variable band width' method, the return probability and the spectral method (see [26] for more details). In the case of the 'variable band width' method, the following quantity is computed :

$$< Zmax(r) >_{r_o} = < \max\{z(r')\}_{r_0<r'<r_o+r} - \min\{z(r')\}_{r_o<r'<r_o+r} >_{r_o} \propto r^\zeta \quad (1)$$

where $r$ is the width of the window. $Zmax(r)$ is the difference between the maximum and the minimum heights $z$ within this window, averaged over all possible origins $r_o$ of the window belonging to the profile.

On the other hand, $\zeta$ can be determined from the scaling of the return probability or of the power spectrum. In the former method, the probability $P(r)$ that the height $z$ goes back to its initial value $z(r=0)$ is computed. We seek for the points $r_i$ which correspond to the same height $z$, and build up the histogram of the differences $r = r_i - r_j$, averaging over all accessible values of $z$ ; it was shown [27, 28, 26] that the averaged histogram $P(r)$ scales in the following way:

$$P(r) \propto r^{-\zeta} \quad (2)$$

In the last method, the power spectrum $S(k)$ of the profile [27] is computed:

$$S(k) \propto k^{-1-2\zeta} \quad (3)$$

where $k$ is the wave vector.

The results of the 'variable band width' method, being less noisy, are analysed first.

In the case of the AFM1/1 zone, three power-law regimes can be observed on the plot of Fig. 2. The roughness exponent which can be determined at short lengthscales is close to 1, which is characteristic of a flat surface. This regime will be discussed in the following.

At larger lengthscales, two regimes can be observed, which correspond to the "quasi-static" (with $\zeta_{QS} \simeq 0.5$) and "dynamic" (with $\zeta \simeq 0.8$) regimes already observed [8] on this material. As in ref.[8], the simplest form of the crossover function is chosen, i.e. $Zmax(r)$ is fitted with the sum of two power laws:

$$Zmax(r) = a0 * ((\frac{r}{\xi_{QS}})^{0.5} + (\frac{r}{\xi_{QS}})^{0.84}) \quad (4)$$

which allows to define the crossover length $\xi_{QS}$. Exponent $\zeta_{QS} \simeq 0.5$ is measured for lengthscales approximately ranging between 1nm and 1$\mu$m, i.e. on roughly three decades. Finally, at lengthscales larger than $\xi_{QS} \simeq 1\mu$m, the "universal" exponent $\zeta \simeq 0.8$ is recovered.

In the case of AFM1/2 (see Fig. 3), the same behaviour is observed. As could be expected, $\xi_{QS}$ is the same as in the AFM1/1 zone. However, the crossover length separating the "flat" and the quasi-static regimes is significantly larger (ranging from 50 nm to 0.1 $\mu$m) than in the previous case. This discrepancy will be discussed in the following.

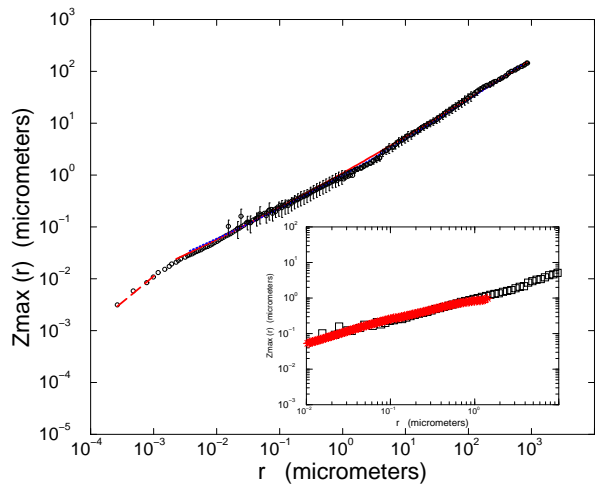

Fig. 2. Regions SEM1 and AFM1/1. $Zmax(r)$ is plotted versus $r$ on a log-log plot. Note that the experimental points obtained with the two techniques gently collapse on the same curve (the region of overlap of the two techniques extending approximately from 10 nm to 1$\mu$m). Three power law regimes can be observed. The slope 1 at the origin is indicated by a dotted line.
Two non linear curve fits of the results are proposed:
— $Zmax(r) = a0((\frac{r}{\xi_{QS}})^{0.5} + (\frac{r}{\xi_{QS}})^{0.78})$ with $a0 = 0.12$ and $\xi_{QS} = 0.1\mu$m.
— — — $Zmax(r) = a0((\frac{r}{\xi_{QS}})^{0.5} + (\frac{r}{\xi_{QS}})^{0.84})$ with $a0 = 0.49$ and $\xi_{QS} = 1\mu$m.
In each case, error bars are estimated from the scattering of experimental results relative to the various micrographs or profiles analysed.
Insert: Region of overlap between AFM ($\star$) and SEM ().

In the case of AFM2 (see Fig. 4), a shrink of the intermediate lengthscales regime is observed. If $Zmax(r)$ is fitted as previously (Eq. (4)), $\xi_{QS}$ is shown to decrease down to 5 nm. However, a slightly better fit is proposed with only one power law (see Fig. 4), with exponent $\zeta = 0.78$. The difference between the two values of $\zeta$ which both fit the experimental data gives an idea of the error bars to be expected on these exponents. For the first time to our knowledge, the universal exponent $\zeta \simeq 0.8$ is observed on *five and a half decades* of lengthscales .

It can be seen in Fig. 2 and 3 that a fit with exponents $\zeta_{QS} = 0.5$ and $\zeta = 0.78$ gives also excellent results in zones SEM1/AFM1/1 and SEM1/AFM1/2, leading then to a crossover length $\xi_{QS} \simeq 0.1\mu$m.

The return probability and the power spectrum analysis give results which are perfectly compatible (see Fig. 5 and 6) with both choices ($\zeta = 0.78$ and $\zeta = 0.84$). In the case of the power spectrum, the cross-over wave vector $k^*_{QS}$ is related to $\xi_{QS}$ by the relation $k^*_{QS}=2\pi/\xi_{QS}$. This leads (region AFM1) to $k^*_{QS} \simeq 6\mu m^{-1}$ for $\xi_{QS}=1\mu m$, and to $k^*_{QS} \simeq 60\mu m^{-1}$ for $\xi_{QS} = 0.1\mu m$, actually corresponding to the cut-offs between a power-law regime with either exponent -2.68 ($-1 - 2\zeta$ with $\zeta = 0.84$, see Eq. (3), Fig. 6), or exponent -2.56 ($-1 - 2\zeta$ with $\zeta = 0.78$, see Fig. 6), and a large wave vectors power law decrease with exponent $-2 = -1 - 2\zeta$ with $\zeta_{QS} = 0.5$. Note that for the computation of the power spectra, only those of the SEM profiles containing 6 000 and 7 000 points (magnification x100) have been used, since profiles with 1 024 points lead to very noisy spectra.

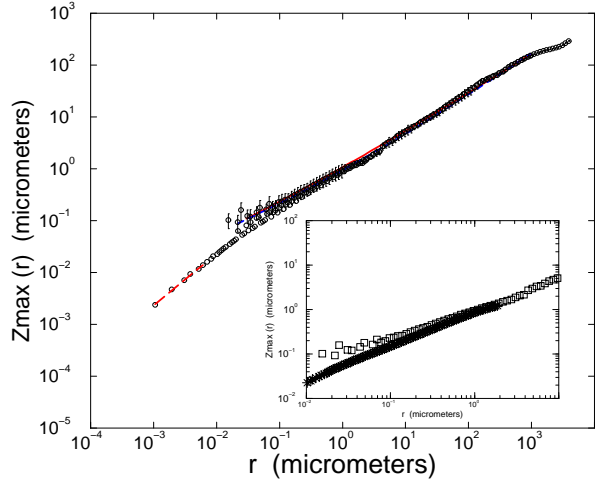

Fig. 3. Regions SEM1 and AFM1/2. $Zmax(r)$ is plotted versus $r$ on a log-log plot. The slope 1 at the origin is indicated by a dotted line.
— $Zmax(r) = a0((\frac{r}{\xi_{QS}})^{0.5} + (\frac{r}{\xi_{QS}})^{0.78})$ with $a0 = 0.11$ and $\xi_{QS} = 0.1 \mu$m.
− − − $Zmax(r) = a0((\frac{r}{\xi_{QS}})^{0.5} + (\frac{r}{\xi_{QS}})^{0.84})$ with $a0 = 0.47$ and $\xi_{QS} = 1 \mu$m.
Insert: Region of overlap between AFM ($\star$) and SEM().

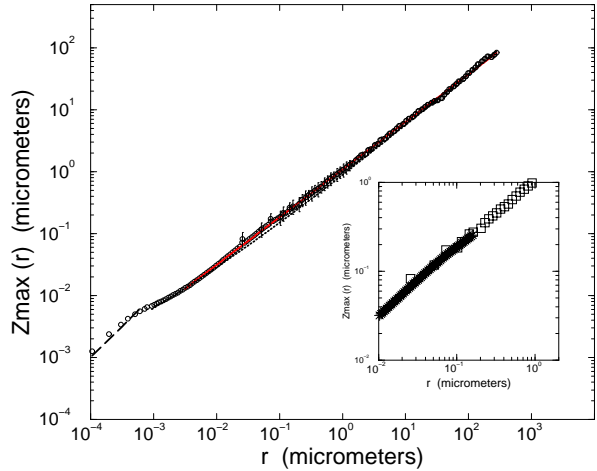

Fig. 4. Regions SEM2 and AFM2. $Zmax(r)$ is plotted versus $r$ on a log-log plot. The slope 1 at the origin is indicated by a dotted line.
Two fits of the data are proposed:
— $Zmax(r) = a0 * r^{0.78}$ with $a0 = 0.08$.
− − − $Zmax(r) = a0((\frac{r}{\xi_{QS}})^{0.5} + (\frac{r}{\xi_{QS}})^{0.84})$ with $a0 = 10$ and $\xi_{QS} = 5nm$.
Note that the "dynamic" regime extends over *five and a half* decades of lengthscales: 5nm-1mm.
Insert: Region of overlap between AFM ($\star$) and SEM().

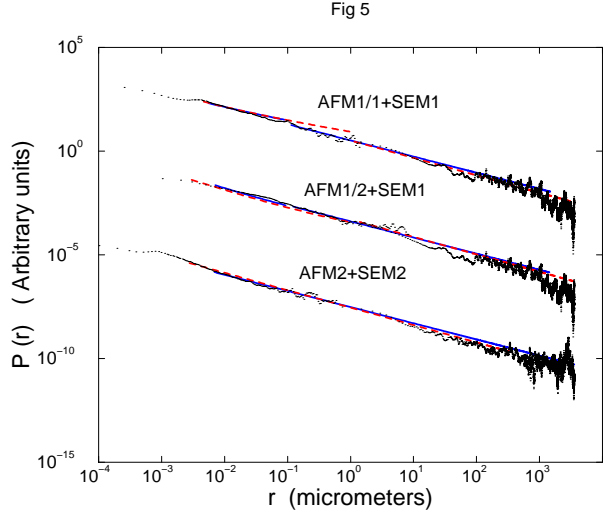

Fig. 5. Return probability for SEM1+AFM1/1, SEM1+AFM1/2 and SEM2+AFM2. Arbitray units for $P$ are chosen in order that the three curves can be put on the same plot.
SEM1+AFM1/1:
– $P(r) = a0(\frac{0.005}{r} + (\frac{0.005}{r})^{0.5})$ for $r < 0.1\mu$m, and $P(r) = \frac{a1}{r^{0.78}}$ for $r > 0.1\mu$m
– – – $P(r) = a0(\frac{0.005}{r} + (\frac{0.005}{r})^{0.5})$ for $r < 1\mu$m, and $P(r) = \frac{a1}{r^{0.84}}$ for $r > 1\mu$m
SEM1+AFM1/2:
– $P(r) = a0(\frac{0.18}{r} + (\frac{0.18}{r})^{0.5})$ for $r < 0.1\mu$m and $P(r) = \frac{a1}{r^{0.78}}$ for $r > 0.1\mu$m
– – – $P(r) = a0(\frac{0.18}{r} + (\frac{0.18}{r})^{0.5})$ for $r < 1\mu$m and $P(r) = \frac{a1}{r^{0.84}}$ for $r > 1\mu$m.
SEM2+AFM2:
– $P(r) = \frac{a1}{r^{0.78}}$
– – – $P(r) = a0(\frac{0.18}{r} + (\frac{0.18}{r})^{0.5})$ for $r < 50$nm and $P(r) = \frac{a1}{r^{0.84}}$ for $r > 50$nm.

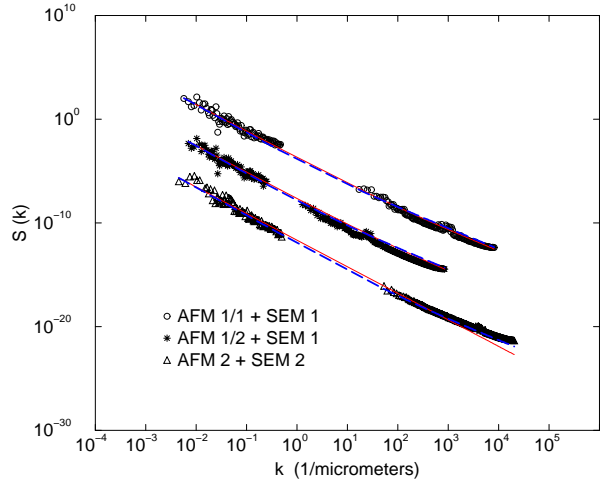

Fig. 6. Power spectrum for SEM1+AFM1/1, SEM1+AFM1/2 and SEM2+AFM2. Arbitray units for $S(k)$ are chosen in order that the three curves can be put on the same plot.
◯ SEM1+AFM1/1:
— $S(k) = a0((\frac{60}{k})^{2.56} + (\frac{60}{k})^{2})$
— — — $S(k) = a0((\frac{6}{k})^{2.68} + (\frac{6}{k})^{2})$
⋆ SEM1+AFM1/2:
— $S(k) = a0((\frac{60}{k})^{2.56} + (\frac{60}{k})^{2})$
— — — $S(k) = a0((\frac{6}{k})^{2.68} + (\frac{6}{k})^{2})$
△ SEM2+AFM2:
— $S(k) = \frac{a0}{k^{2.56}}$
— — — $S(k) = a0((\frac{1200.}{k})^{2.68} + (\frac{1200.}{k})^{2})$.

## III. DISCUSSION

The above results confirm those obtained in [8], where the intermedi-ate and large lengthscales regimes were interpreted respectively as "quasi-static" and "dynamic". They are also in agreement with the experimental results obtained by Milman and coworkers [10, 11] and by Mc Anulty *et al.* [12], and with previously quoted theoretical models [13-17,18-24]. However, the "moving line" models [13-17] predict the short lengthscales regimes to be independant of the crack velocity[29]. This is clear for the first series of experiments [8], and also compatible with the results presented here.

Finally, it has to be noted that the "quasi-static" regime has only been observed on metallic materials [30]. Hence, plasticity might be an important factor for the onset of this regime, either because different fracture mechanisms are indeed involved within the plastic zone, or because plastic dissipation may slow down crack propagation at small lengthscales.

On the other hand, quantitative experiments for fracture in fatigue, where the crack velocity and the load are measured should allow to clarify the meaning of this regime and are currently being performed. In particular, it is not clear from this first set of experiments that the crack velocity is the relevant parameter to be taken into account. Since fracture does not occur at constant load, it is not obvious that region SEM1 corresponds to a smaller stress intensity factor than region SEM2, and our attempts to perform measurements of the load and of the crack velocity in that region failed, because fracture occurs too quickly (approximately in 200 ms, with a sharp variation of the velocity within 10ms). However, it can be noted that an upper bound of the maximum crack velocity could be estimated to be less than 1% of the speed of sound in the material: hence, no inertial effect has to be taken into account.

It would also be of great interest to relate $\xi_{QS}$ to characteristic lengths of the microstructure. Note that $\xi_{QS}$ obtained for AFM1/1 is of the order of the average thickness of the $\alpha_2$ needles. However, in a previous study of the same material [8], a cross-over length of $10\mu$m could be determined for fracture in fatigue. The distribution of the stresses felt by the crack front during its propagation depending both on the microstructural disorder and on the loading conditions, it is expected that $\xi_{QS}$ is linked both to the microstructure, and to the local stress intensity factor. Further experiments on an aluminium alloy will also be performed to help clarifying that point.

Finally, the "flat" regime which should exist at shortest lengthscales is mixed with a non-physical signal linked to the experimental limitations of our AFM. For a given scanning rate, an increase of the number of points registered along a profile of constant length requires an increase of the acqui-sition frequency. In order to avoid any mechanical resonance of the piezo-electric actuator or of the cantilever, the highest frequencies are limited to a few tens of kHz, which imposes a cut-off at short distances. This crossover is then expected to be higher for AFM1/2 than for AFM1/1, which is effectively the case. However, the existence of this non-physical "flat" domain slightly influences the statistical characteristics of the "quasi-static" regime at small lengthscales. As a consequence, AFM1/2 is more "polluted" than AFM1/1. Note also that the importance of this artefact may be more or less important depending on the statistical methods used.

In conclusion, it has been shown that the results obtained with AFM are *quantitatively* compatible with those obtained with standard SEM. The simultaneous use of both techniques has allowed, for the first time, an observation of the "universal" fracture regime, characterised by a roughness index $\zeta \simeq 0.8$, over roughly *six decades* of lengthscales. At small enough lengthscales, a "quasi-static" regime spreading over three decades has been observed, allowing a good determination of its roughness exponent $\zeta_{QS} \simeq 0.5$. The cross-over length $\xi_{QS}$ between the two regimes decreases in regions farther from the initial notch.

### Acknowledgments

Fracture experiments were achieved in collaboration with G. Marcon. We are grateful to M. Thomas for sharing with us his vast knowledge on the microstructures of $Ti_3Al$-based alloys. Enlightening discussions with J.-P. Bouchaud, J.-L. Chaboche, G. Lapasset, A. Pineau and S. Roux are also acknowledged. Finally, we are indebted to Pr B. Mandelbrot for his encouragements.

1. *

References

[1] B.B. Mandelbrot, D.E. Passoja, A.J. Paullay, Nature **308** 721 (1984).

[2] A. Imre, T. Pajkossy, L. Nyikos, Acta Metall. Mater. **40**,8,1819 (1992).

[3] J. Schmittbuhl, S. Gentier, S. Roux, Geophys. Research Lett. **20**, 8, 639 (1993).

[4] E. Bouchaud, G. Lapasset, J. Planès, S. Navéos, Phys.rev. B **48**, 2917 (1993).

[5] J. Planès, E. Bouchaud, G. Lapasset, Fractals **1**, 1059 (1993).

[6] E. Bouchaud, G. Lapasset, J. Planès, Europhys. Lett. **13**,73 (1990).

[7] J.J. Mecholsky, D.E. Passoja, K.S. Feinberg-Ringel, J. Am. Ceram. Soc. **72**,60 (1989).

[8] E. Bouchaud, S. Navéos, J.Phys. I France **5**, 547 (1995).

[9] E. Bouchaud, J.P. Bouchaud, Phys. Rev. B **50**, 17752 (1994).

[10] V.Y. Milman, R. Blumenfeld, N.A. Stelmashenko, R.C. Ball, Phys. Rev. Lett. **71** 204 (1993).

[11] V.Y. Milman, N.A. Stelmashenko, R. Blumenfeld, preprint (1994)

[12] P. McAnulty, L.V. Meisel, P.J. Cote, Phys. Rev. A **46**, 3523 (1992).

[13] J.-P. Bouchaud, E. Bouchaud, G. Lapasset, J. Planès Phys. Rev. Lett. **71**, 2240 (1993) ; E. Bouchaud, J.-P. Bouchaud, G. Lapasset, J. Planès, Fractals, **1**, 1051 (1993).

[14] D. Ertas, M. Kardar Phys.Rev.Lett. **69**,929 (1992).

[15] D. Ertas, M. Kardar Phys.Rev, E **48** 1228 (1993).

[16] D. Ertas, M. Kardar Phys.Rev.Lett. **73**,1703 (1994).

[17] D. Ertas, M. Kardar, preprint (1995).

[18] S. Roux, D. François, Scripta Metall. **25**, 1092 (1991).

[19] A. Hansen, E.L. Hinrichsen, S. Roux, Phys. Rev. Lett., **66**, 2476 (1991).

[20] M. Kardar, Nucl. Phys. **B290**, [FS20], 582 (1987).

[21] M. Mézard, G. Parisi, J. Phys. France I **1**, 809 (1991).

[22] T. Halpin-Healy, Phys. Rev. A **42**, 711 (1990).

[23] G. Batrouni, S. Roux, preprint (1995).

[24] A.A. Middleton, preprint (1995).

[25] J. Schmittbuhl, S. Roux, J.P. Vilotte, K.J. Maloy, Phys. Rev. Lett. **74**, 1787 (1995).

[26] J. Schmittbuhl, J.P. Vilotte, S. Roux, Phys. Rev. E **51**, 131 (1995).

[27] M.B. Isichenko, Review of Modern Physics **64**, 961 (1992).

[28] J. Feder, *Fractals*, Plenum Press (New-York), (1988).

[29] S. Roux, private communication (1995).

[30] F. Plouraboué, K. W. Winkler, L. Petitjean, J.-P. Hulin, S. Roux, preprint (1995).